\documentclass[11pt]{article}
\usepackage{amsmath,amsfonts,amssymb}
\usepackage{graphicx}
\usepackage[margin=1in]{geometry}
\usepackage{authblk}
\usepackage{cite}
\usepackage{hyperref}
\usepackage{microtype}
\hypersetup{colorlinks=true,citecolor=blue,linkcolor=blue,urlcolor=blue}
\title{DOLPHIN (Device of Line Photometry in Nasmyth): A Switchable Multiband Photometer for the ISAS 1.3-m Telescope}

\author[a]{Kohei Miyakawa}
\author[a]{Hirokazu Kataza}
\author[b,c,d]{Teruyuki Hirano}
\author[a]{Hajime Kawahara}
\author[a]{Shotaro Tada}
\author[c,d]{Takehiko Wada}

\affil[a]{Institute of Space and Astronautical Science, Japan Aerospace Exploration Agency, 3-1-1 Yoshinodai, Chuo-ku, Sagamihara City, Kanagawa 252-5210, Japan}
\affil[b]{National Astronomical Observatory of Japan, 2-21-1 Osawa, Mitaka, Tokyo 181-8588, Japan}
\affil[c]{Astrobiology Center, 2-21-1 Osawa, Mitaka, Tokyo 181-8588, Japan}
\affil[d]{Department of Astronomical Science, School of Physical Sciences, The Graduate University for Advanced Studies (SOKENDAI), 2-21-1, Osawa, Mitaka, Tokyo, 181-8588, Japan}

\date{}
\begin{document} 
\maketitle

\begin{abstract}
We present the design, construction, and commissioning of DOLPHIN
(Device of Line Photometry in Nasmyth), a compact instrument for switchable
broad- and narrow-band photometry at the Nasmyth focus of the ISAS
1.3-m telescope. DOLPHIN was developed as a cost-conscious platform for
differential measurements of selected spectral features with modest-aperture
telescopes. A motorized eight-position filter wheel provides access to
Johnson--Bessell $B$, $V$, and $R$ bands and narrow-band Na D and
H$\alpha$ filters. All filters share a common optical path and CMOS detector,
and adjacent positions can be switched in approximately 0.8~s. The optical
and mechanical systems use commercially available components and provide an
approximately 10-arcmin field of view.

Laboratory characterization shows that the detector response is linear to
within 0.1\% below 10,000~ADU, with a conversion factor of
0.81~e$^-$/ADU, read noise of 9~e$^-$, and pixel-response non-uniformity of
0.37\%. We also measure the in-system transmission profiles of all filters
and identify a broader-than-specified Na D passband. Commissioning
observations of HD~189733 are used as an engineering validation of
filter-sequence operation, image calibration, and relative photometric
stability. The commissioning sequence used 1-s exposures in $B$, $V$, and
$R$ and 20-s exposures in Na D and H$\alpha$. Under these settings, the
unbinned RMS ranges from 0.34\% in Na D to 0.94\% in $B$; after eight-point
binning, the Na D/$R$ ratio reaches 0.16\%.
Positive inter-band correlations that increase with bin size demonstrate
the removal of part of the low-frequency common-mode noise. Recovery of the
known broadband transit provides an end-to-end validation of the observing
and reduction system. These results establish DOLPHIN as a practical
platform for differential narrow-band monitoring and define the
instrumental limitations to be addressed in future observations.
\end{abstract}


\medskip
\noindent\textbf{Corresponding author:} Kohei Miyakawa
(\href{mailto:miyakawa@ir.isas.jaxa.jp}{miyakawa@ir.isas.jaxa.jp})

\section{Introduction}
\label{sect:intro}  
Time-series measurements of selected spectral features are used in a wide
range of astronomical applications, including studies of stellar activity
and the wavelength dependence of exoplanet transits. High-resolution
spectroscopy provides detailed line profiles, but it requires a dispersive
optical system, stable calibration, and typically a large-aperture telescope.
For programs that require repeated measurements of only a small number of
preselected wavelength intervals, imaging through narrow-band filters
offers a simpler alternative.

Several instruments have demonstrated the value of filter-based differential
measurements, although the closest narrow-band implementations have
generally relied on large facilities. The tunable-filter mode of OSIRIS on
the 10.4-m Gran Telescopio Canarias was used for nearly simultaneous
narrow-band transit photometry,\cite{Colon2010} while a 0.635-nm
narrow-band filter installed in WIRC on the 5.1-m Hale Telescope constrained
metastable-helium absorption.\cite{Vissapragada2020} HIRAX, likewise designed
for the 5.1-m Hale Telescope, extends this concept to three simultaneous
0.03-nm bands around the Na D doublet.\cite{Baker2024} These examples
establish the feasibility of differential narrow-band measurements,
but also illustrate their reliance to date on 5--10-m-class facilities.

The instrumental challenge is not simply to obtain sufficient counts in a
narrow passband. Variations in atmospheric transparency, image position,
seeing, and detector response can exceed the differential signal of
interest. A useful narrow-band photometer must therefore measure a spectral
feature together with one or more continuum references on a cadence short
enough to track these variations. Simultaneous multi-channel instruments
can provide this information but require beam splitters, separate optical
trains, and cross-calibration among detectors. A sequential filter system
cannot provide strictly simultaneous measurements, but it places every band
on the same detector and optical path and can be implemented with
substantially less optical and mechanical complexity.

We developed DOLPHIN (Device of Line Photometry in Nasmyth) to evaluate this
single-channel, rapidly switchable approach on the ISAS 1.3-m telescope and
to test whether selected spectral-line monitoring can be extended to a
modest-aperture facility. The instrument was designed around four engineering
priorities: a shared optical path for all passbands, short and repeatable
filter changes, compatibility with both broadband and narrow-band
filters, and construction from commercially available components wherever
practical. The resulting system is compact, accessible at the Nasmyth
platform, and can be modified or replicated without specialized dispersive
optics. Although the smaller collecting area limits the accessible targets,
this architecture is efficient in observing resources and instrument
complexity for bright-star programs: it avoids competition for
large-telescope time as well as the construction and cross-calibration of
multiple optical channels, enabling repeated line-monitoring campaigns with
a 1-m-class telescope.

This paper is primarily an instrument-development and commissioning report.
We describe the design trade-offs, optical and mechanical implementation,
filter set, detector characteristics, and control strategy. We then present
laboratory measurements of detector and in-system filter performance,
followed by an on-sky engineering evaluation. HD~189733 was selected because
its deep, well-characterized transit provides a convenient end-to-end test
signal. The observations are used to quantify photometric precision,
time-correlated noise, inter-band common-mode rejection, and recovery of a
known broadband signal; they are not intended as a definitive measurement
of the planetary atmosphere.

\section{Instrument Design}
\subsection{Design Policy}

DOLPHIN was developed to provide observational constraints on spectral-line variability associated with exoplanet atmospheres and stellar activity using a relatively low-cost instrument. 
Rather than employing a conventional spectrograph, DOLPHIN adopts a switchable filter system that performs multi-wavelength observations by sequentially acquiring images through different filters.

Compared with simultaneous multi-band systems based on dichroic beam splitters, the filter-switching approach sacrifices temporal resolution but offers several important advantages. 
First, spectral-line and continuum measurements can be obtained using filters with nearly identical bandpasses and optical characteristics, enabling precise differential photometry between a target absorption line and its adjacent continuum. 
Second, all wavelength bands share the same optical path and detector system, allowing instrumental systematics to be largely common among the measurements and thereby reducing differential calibration errors. 
Third, the optical design is significantly simpler, more compact, and less expensive than multi-channel imaging systems, facilitating deployment on small- and medium-sized telescopes.

To maximize reproducibility and minimize development costs, the instrument was designed using commercially available off-the-shelf components whenever possible. 
This design philosophy enables straightforward replication of the instrument for future installations at additional observatories.

The installation site of the ISAS/JAXA 1.3-m telescope is located in an urban environment. Several artificial light sources, including illuminated sports facilities operating at night, are situated within approximately 500 m of the observatory. 
Consequently, suppression of stray light was identified as a key design requirement. 
The optical and mechanical structures of DOLPHIN therefore incorporate dedicated baffling and shielding measures to minimize contamination from off-axis artificial light sources and maintain photometric stability during observations.


\subsection{Optics and Mechanics}

\begin{figure}
\begin{center}
\includegraphics[width=0.98\linewidth]{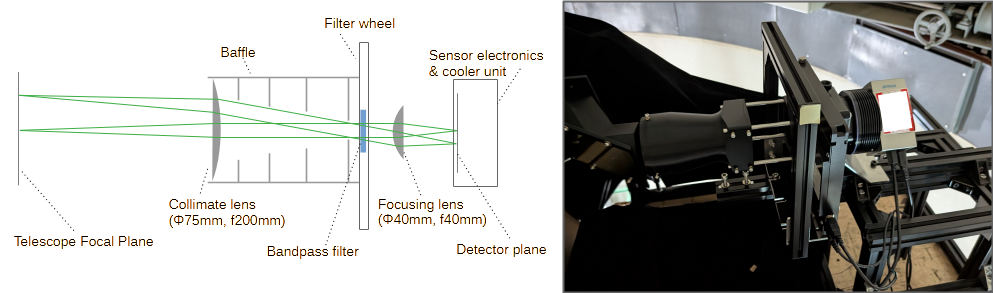}  
\\
\end{center}
\caption 
{ \label{fig:optical_design}
Optical layout of DOLPHIN (left) and the assembled instrument installed at
the Nasmyth focus (right).} 
\end{figure} 

\begin{table}[ht]
\caption{Major optical and mechanical components of DOLPHIN.} 
\label{tab:components}
\begin{center}       
\small
\begin{tabular}{|l|l|l|l|} 
\hline
\rule[-1ex]{0pt}{3.5ex}  Component & Manufacturer&Model& Specification \\
\hline\hline
\rule[-1ex]{0pt}{3.5ex}  Collimator lens & Edmund Optics & \#86-913 & $f=200$ mm, $\phi=75$ mm  \\
\rule[-1ex]{0pt}{3.5ex}  Focusing lens & Edmund Optics & \#48-663 & $f=40$ mm, $\phi=40$ mm  \\
\rule[-1ex]{0pt}{3.5ex}  Motorized filter wheel & Edmund Optics & \#23-646, \#59-770 & Eight 1-inch slots\\
\rule[-1ex]{0pt}{3.5ex}  Detector unit & Bitran & CS-74M & APS-C CMOS sensor\\
\hline\hline 
\end{tabular}
\end{center}
\end{table} 

The optical layout and a photograph of the assembled DOLPHIN instrument are
shown in Fig.~\ref{fig:optical_design},
and the major optical and mechanical components are summarized in Table~\ref{tab:components}.

Light from the telescope first forms an image at the telescope focal plane and subsequently enters the collimator lens of DOLPHIN. 
The collimated beam then passes through a series of stray-light suppression baffles before reaching the filter wheel, which is positioned near a re-imaged pupil plane. 
Locating the filters close to the pupil reduces the required filter aperture and minimizes wavelength shifts associated with non-normal incidence. 
After transmission through the selected filter, the beam is focused onto the detector by the focusing lens.

The focal lengths of the collimator and focusing lenses were selected to be 200 mm and 40 mm, respectively. 
These values were determined by considering several practical constraints, including the overall instrument dimensions, the required field of view, and mechanical compatibility with the telescope optics and image rotator. 
The resulting optical configuration provides a field of view of approximately 10 arcmin in diameter, which is primarily determined by the clear aperture of the collimator lens.
Since DOLPHIN is optimized for photometry rather than imaging, stringent aberration correction is not required. 
Furthermore, the typical seeing conditions at the observatory are substantially larger than the expected residual optical aberrations. 
Consequently, the optical system is constructed from spherical lenses, providing a cost-effective and compact design while maintaining sufficient image quality for the intended scientific observations.

\begin{table}[ht]
\caption{Filters currently installed in the DOLPHIN filter wheel. CWL and FWHM denote the catalog central wavelength and full width at half maximum, respectively.} 
\label{tab:filters}
\begin{center}       
\begin{tabular}{|c|c|c|c|c|} 
\hline
\rule[-1ex]{0pt}{3.5ex}  Index & Band & Manufacturer & Model & CWL, FWHM [nm] \\
\hline\hline
\rule[-1ex]{0pt}{3.5ex}  1 & Dark & & &  \\
\rule[-1ex]{0pt}{3.5ex}  2 & $B$ & Edmund Optics & \#21-122 & 440.0, 100.0 \\
\rule[-1ex]{0pt}{3.5ex}  3 & Na D & Alluxa & 589-1 OD4 & 589.0, $<$1.2 \\
\rule[-1ex]{0pt}{3.5ex}  4 & $R$ & Edmund Optics & \#21-124 & 630.0, 120.0 \\
\rule[-1ex]{0pt}{3.5ex}  5 & H$\alpha$ & Alluxa & 656.3-1 OD4& 656.3, $<$1.2 \\
\rule[-1ex]{0pt}{3.5ex}  6 & empty & & &  \\
\rule[-1ex]{0pt}{3.5ex}  7 & $V$ & Edmund Optics & \#21-123 & 520, 90.0  \\
\rule[-1ex]{0pt}{3.5ex}  8 & empty & & &  \\
\hline\hline 
\end{tabular}
\end{center}
\end{table} 

DOLPHIN employs a motorized filter wheel that enables rapid switching among multiple observing bands. The filter wheel is controlled through a USB interface and utilizes bi-directional positioning,
allowing efficient selection of arbitrary filter positions. 
The switching time between adjacent filter positions is approximately 0.8 s.
The wheel accommodates up to eight 1-inch diameter filters, providing substantial flexibility for multi-band observations. 
This configuration allows the instrument to alternate between continuum and spectral-line filters with minimal overhead, 
enabling quasi-simultaneous differential photometry. 

The current filter set installed in DOLPHIN is summarized in Table~\ref{tab:filters}.
Narrow-band filters centered on the Na D and H$\alpha$ lines are included as the primary science filters for studies of exoplanet atmospheres and stellar activity. 
Johnson--Bessell $V$ and $R$ filters are used as continuum references, while a $B$ filter is included to monitor broadband flux variations.
The narrow-band filters were purchased from the Alluxa catalog, whereas the broadband filters are commercially available Johnson--Bessell filters from Edmund Optics.
One filter-wheel position is masked with a black sealing and is used for dark-frame acquisition.

DOLPHIN employs a CS-74M camera manufactured by Bitran. The CS-74M integrates the image sensor, readout electronics, and thermoelectric cooling system into a single compact unit.
The detector is a Sony IMX571BLR-J CMOS sensor with an APS-C format (23.5 $\times$ 15.7 mm), consisting of 6244 $\times$ 4168 pixels with a pixel size of 3.76~${\rm\mu m}$. 
Owing to the available telescope field of view and mechanical constraints of the instrument, the full sensor area cannot be uniformly illuminated. 
The full 6244 $\times$ 4168 pixel array is read out for every exposure, but
the effective science region is restricted to the central 3460 $\times$
3460 pixels, which encompass the optical image circle.
All observations are acquired in the 16-bit readout mode. The detector is cooled using a thermoelectric (Peltier) cooling system, typically operated at approximately 20 K below the ambient temperature.

\subsection{ISAS 1.3-m Telescope}
\begin{figure}
\begin{center}
\includegraphics[height=7cm]{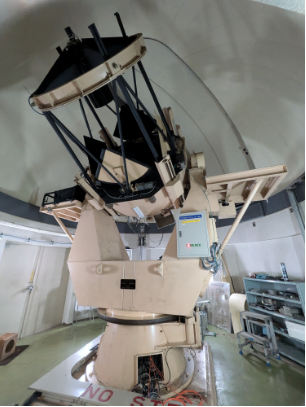}
\end{center}
\caption{\label{fig:isas_telescope}
The ISAS/JAXA 1.3-m telescope at the Sagamihara campus.}
\end{figure}

The ISAS/JAXA 1.3-m telescope is shown in
Fig.~\ref{fig:isas_telescope}. Since no dedicated technical publication
describing the telescope is available, a brief overview is provided here.
The telescope was constructed in 1988 at ISAS/JAXA as the first alt-azimuth telescope developed in Japan and served as a prototype for the later development of the Subaru Telescope. It is a 1.3-m aperture telescope with an original focal ratio of F/18 and was initially designed for infrared observations.
In the mid-2000s, the primary mirror coating was changed from gold to aluminum, extending the operational wavelength range to visible and near-infrared observations. In 2024, the telescope control system was upgraded as part of an ongoing redevelopment effort aimed at enabling routine scientific observations and expanding the telescope's research capabilities.

As an early demonstration of its renewed scientific capability, the
refurbished telescope contributed transit photometry of XO-3b to a joint
analysis with XMM-Newton and TESS data.\cite{Cilley2026} These observations
did not use DOLPHIN.

DOLPHIN is installed at the western Nasmyth focus of the telescope. The Nasmyth platform provides convenient access to the instrument, facilitating iterative hardware modifications, alignment procedures, and performance optimization during the commissioning phase.

\section{Laboratory Characterization}
\subsection{Filter Transmittance in DOLPHIN}
\begin{figure}
\begin{center}
\begin{tabular}{c}
\includegraphics[height=5cm]{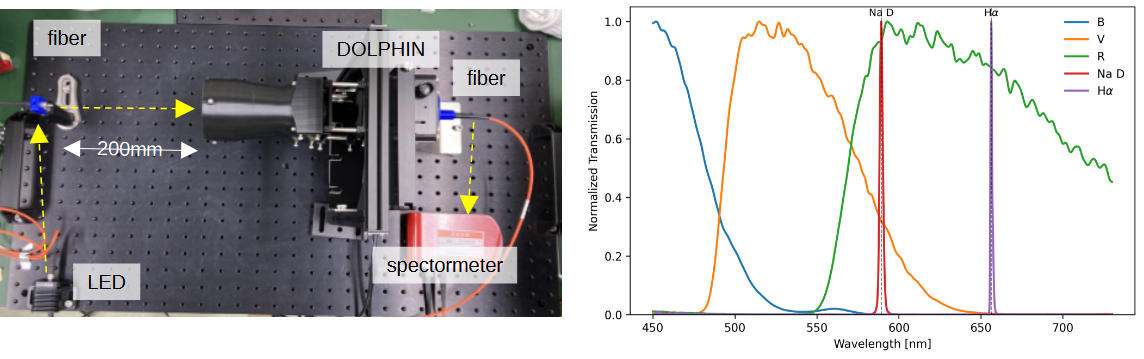}  
\\
(a) \hspace{5.1cm} (b)
\end{tabular}
\end{center}
\caption{\label{fig:trans}
(a) Laboratory setup used to measure the in-system filter transmission.
The yellow arrows indicate the optical path from the fiber-fed broadband
source through DOLPHIN to the spectrometer. (b) Measured transmission curves,
obtained by dividing each filtered spectrum by a reference spectrum acquired
without a filter and then normalizing to the peak of each filter.}
\end{figure} 

\begin{figure}
\begin{center}
\begin{tabular}{c}
\includegraphics[height=6cm]{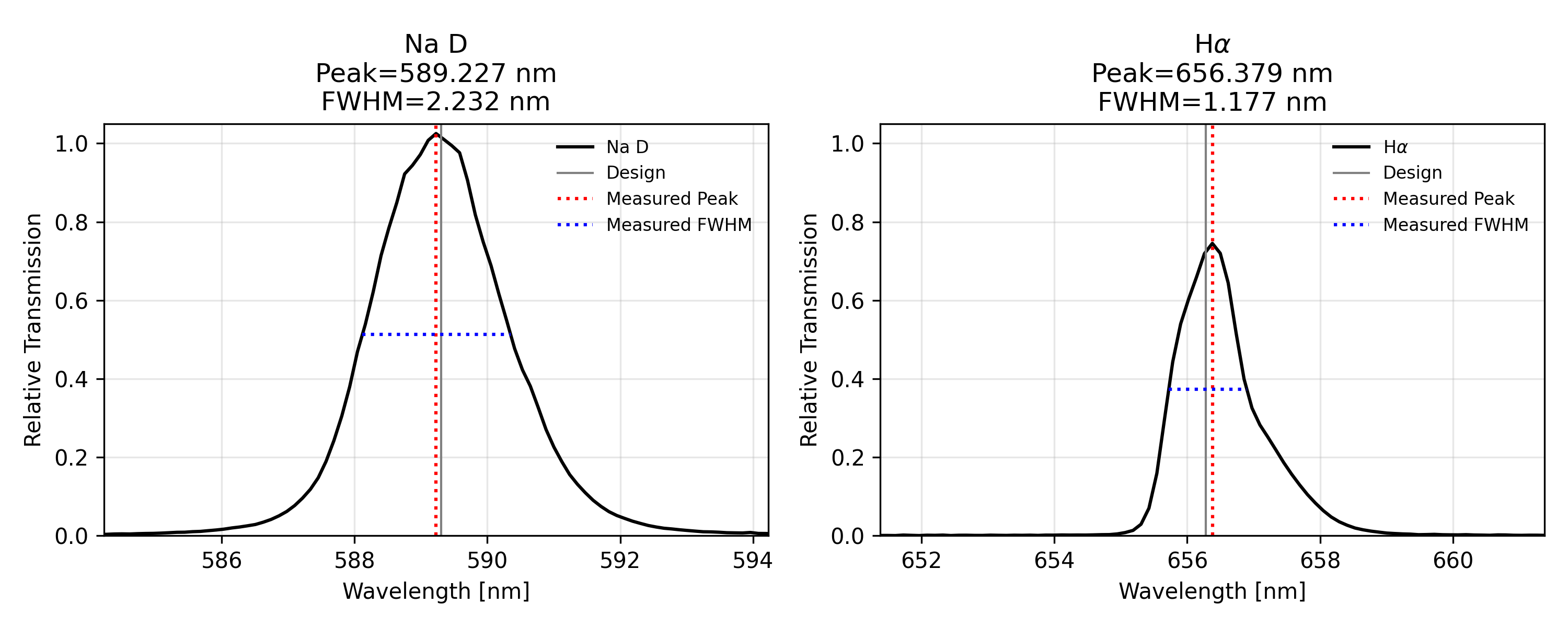}  
\\
\end{tabular}
\end{center}
\caption{\label{fig:narrow}
Measured transmission profiles of the Na D (left) and H$\alpha$ (right)
filters. Red dotted lines mark the measured peak wavelengths, blue dotted
lines show the measured FWHM, and gray lines indicate the catalog central
wavelengths.}
\end{figure} 

The transmission characteristics of the filters installed in DOLPHIN were measured using the setup shown in Figure~\ref{fig:trans}. 
During the measurement, the camera sensor unit of DOLPHIN was temporarily replaced with an optical fiber connected to a spectrometer. 
A fiber-coupled LED source was positioned near the focal plane of the collimator lens to illuminate the optical system. 
The optical path is indicated by the yellow arrows in Figure~\ref{fig:trans}.
The alignment was performed by first adjusting the fiber source so that its image was centered on the DOLPHIN detector. 
The detector was then replaced by the spectrometer fiber input while maintaining the same optical alignment. 
The light source and spectrometer were a Thorlabs MBB1F1 broadband LED source and a Thorlabs CSS100 compact spectrometer, respectively. 
For each filter, the relative transmission spectrum was calculated by
dividing the spectrum acquired with the filter inserted by a reference
spectrum obtained through the same optical path without a filter.
Owing to the spectral coverage of these devices, the transmission measurements were limited to the wavelength range of 450--700 nm.

Figure~\ref{fig:trans}(b) shows the measured transmission curves of the filters installed in DOLPHIN. For clarity, the broadband filter curves were smoothed using a Gaussian filter. 
After division by the no-filter reference, all transmission curves were
normalized to their respective peak values for comparison of their profile
shapes. No significant out-of-band leakage was detected for either of the narrow-band filters within the measured wavelength range.

Figure~\ref{fig:narrow} presents enlarged views of the Na D and H$\alpha$ filters. For both filters, the measured central wavelengths were consistent with the manufacturer specifications within the spectral resolution of the measurement system ($\sim$0.5 nm). 
However, the measured full width at half maximum (FWHM) of the Na D filter was approximately twice the specified value. 
The H$\alpha$ filter exhibited a FWHM of approximately 1.2 nm, while its peak transmission was measured to be about 75\%, which is lower than the catalog specification.

At present, the origin of these discrepancies has not been conclusively identified. Possible contributors include the optical configuration of DOLPHIN, manufacturing tolerances of the filters, and systematic uncertainties in the transmission measurement setup, such as fiber-coupling efficiency. 
Nevertheless, the measured filter characteristics remain adequate for the primary science objectives of DOLPHIN, and these deviations are not expected to have a significant impact on its observational performance.

\subsection{Detector Unit Evaluation}
\begin{figure}
  \begin{center}
    \begin{tabular}{c}
    \includegraphics[height=5.5cm]{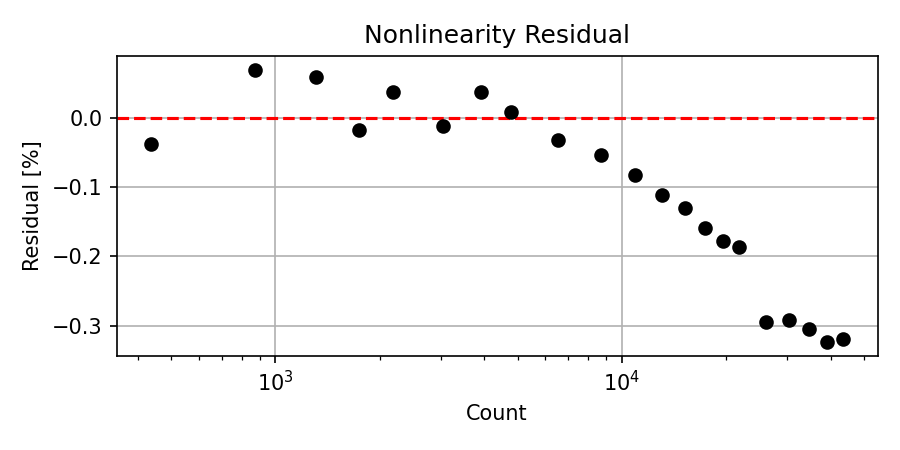}
    \end{tabular}
  \end{center}
  \caption{\label{fig:linearity}
  Residual detector response after subtraction of the best-fitting linear
  relation. The red dashed line marks zero residual.}
\end{figure} 
Laboratory characterization was performed to evaluate the key performance parameters of the CS-74M detector unit used in DOLPHIN. 
The camera employs a Sony IMX571BLR-J CMOS sensor with a specified full-well capacity of 44,000 e$^-$. 
DOLPHIN is operated in the 16-bit readout mode. 
Since the instrument is intended for outdoor operation, the detector temperature was set to 10$^\circ$C, which provides stable thermal control under typical observing conditions.

The detector linearity and conversion factor were measured using a broadband visible LED light source coupled to an integrating sphere through an optical fiber. 
The uniformly illuminated output of the integrating sphere was imaged onto the detector. 
Exposure times were varied from 0.01 s to 1 s to obtain images spanning a wide range of signal levels.

Figure~\ref{fig:linearity} shows the residuals from a linear fit. 
The detector remains linear within 0.1\% below 10,000 ADU, corresponding to
the typical operating range of DOLPHIN. 
The maximum nonlinearity measured over the full test range was approximately
0.3\%.

The conversion factor and readout noise were estimated using the photon transfer method. 
Pairs of images obtained with identical exposure times were subtracted, and the variance of the difference images was analyzed as a function of signal level. 
From this analysis, the conversion factor was determined to be $0.81\pm0.00$ e$^-$/ADU and the readout noise was estimated to be $9.1\pm0.4$ e$^-$.

\begin{figure}
  \begin{center}
    \begin{tabular}{c}
    \includegraphics[width=0.95\linewidth]{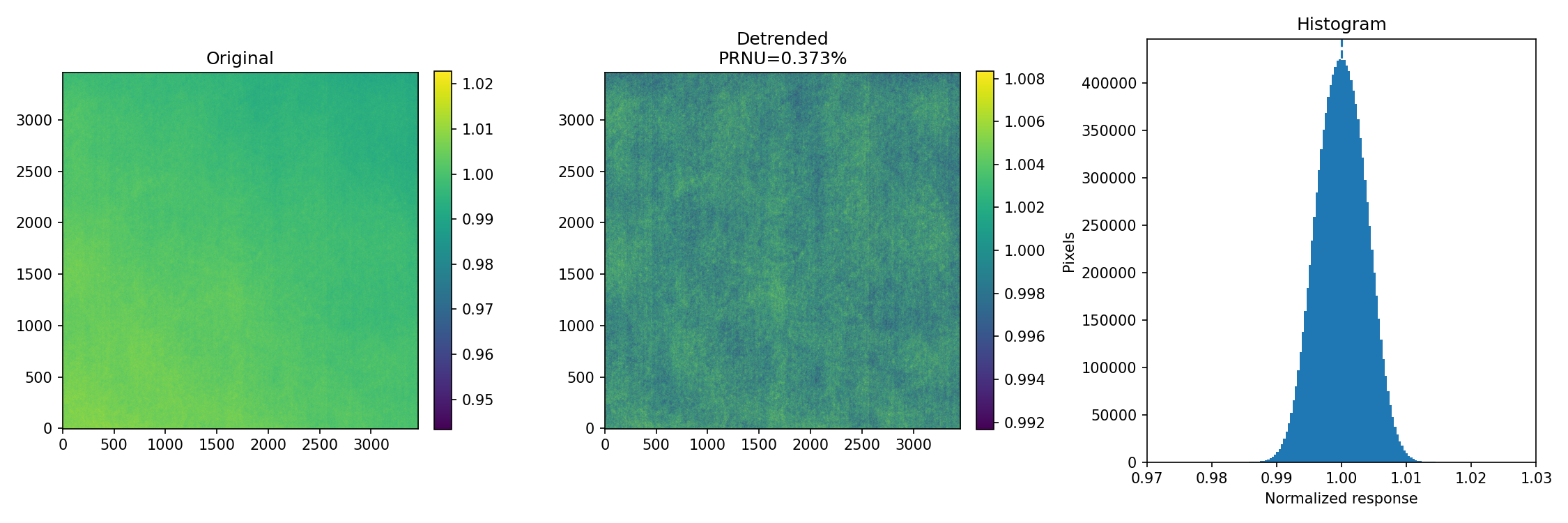}
    \end{tabular}
  \end{center}
  \caption{\label{fig:flat}
  Pixel-response map after dark subtraction and removal of the large-scale
  illumination pattern. The distribution at right shows the normalized
  pixel-response values used to estimate the response non-uniformity.}
\end{figure} 

Flat-field calibration frames were acquired using the same laboratory setup described above. 
The illumination level was adjusted to produce signal levels of approximately 20,000 ADU. 
After dark-current subtraction, the individual frames were median-combined to generate a master flat.

To isolate pixel-to-pixel sensitivity variations, large-scale illumination gradients were removed by applying a Gaussian smoothing filter and dividing the master flat by the resulting low-frequency component. 
The processed flat field is shown in Fig.~\ref{fig:flat}. 
The pixel response non-uniformity was measured to be approximately 0.37\%.
No nonlinearity correction was applied to the flat fields. The maximum
nonlinearity of 0.3\% contributes only about 11~ppm to the measured 0.37\%
pixel-response variation and is negligible for the present analysis.

The dark current was evaluated with the detector installed in DOLPHIN under nominal operating conditions. A dark current of approximately 0.01~e$^{-}$~pixel$^{-1}$~s$^{-1}$ was measured.

\section{On-Sky Performance}
\subsection{Target: HD~189733b}
HD~189733 is a bright ($V\simeq7.7$) K-type dwarf hosting a hot Jupiter
with a deep, well-characterized transit. We selected this system as an
end-to-end commissioning target because its known broadband signal can be
recovered in a single night and its field contains a usable comparison star.

\subsection{Operation Settings}
The commissioning observation of HD~189733b began on 2026 May 18 at
23:56 Japan Standard Time (14:56 UTC). The sky remained clear throughout
the observing sequence.

\begin{table*}[ht]
\caption{Exposure settings, representative count levels, and expected
single-exposure statistical noise for the HD~189733 observations. The sky
count is the mean rate per pixel, and $N_{\rm ap}$ is the representative
number of pixels in the source aperture. The expected noise includes source
photon noise, sky-background noise, and read noise.} 
\label{tab:sequence}
\begin{center}       
\resizebox{\textwidth}{!}{%
\begin{tabular}{|c|c|c|c|c|c|}
\hline
\rule[-1ex]{0pt}{3.5ex} Filter & Exp. time [s] &
Total count [e$^{-}$] & Sky count [e$^{-}$ pixel$^{-1}$ s$^{-1}$] &
$N_{\rm ap}$ [pixel] & Expected noise [\%] \\
\hline\hline
\rule[-1ex]{0pt}{3.5ex}  $R$ & 1 & $1.6 \times 10^6$ & 15 & 8000 & 0.10 \\
\hline
\rule[-1ex]{0pt}{3.5ex}  Na D & 20 & $5.6 \times 10^5$ & 0.6 & 5000 & 0.18 \\
\hline
\rule[-1ex]{0pt}{3.5ex}  $V$ & 1 & $9.6 \times 10^5$ & 18 & 6000 & 0.13 \\
\hline
\rule[-1ex]{0pt}{3.5ex}  H$\alpha$ & 20 & $2.4 \times 10^5$ & 0.1 & 4500 & 0.33 \\
\hline
\rule[-1ex]{0pt}{3.5ex}  $B$ & 1 & $2.8 \times 10^5$ & 8 & 9000 & 0.37 \\
\hline\hline 
\end{tabular}
}
\end{center}
\end{table*} 

The filter sequence adopted for the observations is summarized in Table~\ref{tab:sequence}. 
To measure relative variations of the H$\alpha$ and Na D lines with respect to the continuum, observations in the Johnson--Bessell $V$ and $R$ bands were interleaved with the narrow-band measurements. 
A $B$-band exposure was also included to monitor broadband color variations throughout the observing sequence.

For comparison with the on-sky scatter, the expected single-exposure
statistical noise was calculated as
\begin{equation}
\sigma_{\rm ideal} =
\frac{\sqrt{F+N_{\rm ap}\left(B_{\rm sky}t+\sigma_{\rm read}^{2}\right)}}{F},
\end{equation}
where $F$ is the net source count defined in Eq.~(\ref{eq:net_source}),
$B_{\rm sky}$ is the mean sky count 
per pixel per second, $t$ is the exposure time, and
$\sigma_{\rm read}=9$~e$^-$~pixel$^{-1}$ is the measured read noise. The
representative aperture sizes and resulting values of
$\sigma_{\rm ideal}$ are listed in Table~\ref{tab:sequence}. These
values represent the expected noise from counting statistics and detector
readout alone; they do not include scintillation, transparency variations,
image motion, flat-field residuals, or other time-correlated effects.

Target acquisition, filter-dependent guiding, and focus compensation were
performed automatically during the sequence. Operational details are given
in Appendix~\ref{app:operation}.

\subsection{Analysis Overview}
Science frames were dark-subtracted and flat-fielded before aperture
photometry. We measured single-band stability from detrended light curves
and common-mode rejection from filter-ratio light curves averaged within
each switching cycle. For the signal-recovery test, each ratio light curve
was fitted with the ratio of two passband-specific transit models, including
separate limb-darkening coefficients, rather than with a single transit
model. The full aperture definition, background estimator,
and uncertainty calculation are described in
Appendix~\ref{app:reduction}; the ratio model is defined in
Appendix~\ref{app:signal_recovery}.

\subsection{Results}
\subsubsection{Photometric Precision}
\begin{table}
\centering
\caption{
Photometric precision measured in each filter.
The RMS values are given in percent after binning
the light curves by different factors.
}
\label{tab:precision}
\begin{tabular}{|l|cccc|}
\hline
Filter &
Bin 1 &
Bin 2 &
Bin 4 &
Bin 8 \\
&
(\%) &
(\%) &
(\%) &
(\%) \\
\hline
{\it R}       & 0.50 & 0.36 & 0.29 & 0.24 \\
Na D    & 0.34 & 0.28 & 0.22 & 0.16 \\
{\it V}       & 0.53 & 0.40 & 0.33 & 0.20 \\
H$\alpha$ & 0.59 & 0.43 & 0.33 & 0.25 \\
{\it B}       & 0.94 & 0.70 & 0.53 & 0.40 \\
\hline
\end{tabular}
\end{table}
\begin{figure}
  \begin{center}
    \begin{tabular}{c}
    \includegraphics[height=10.0cm]{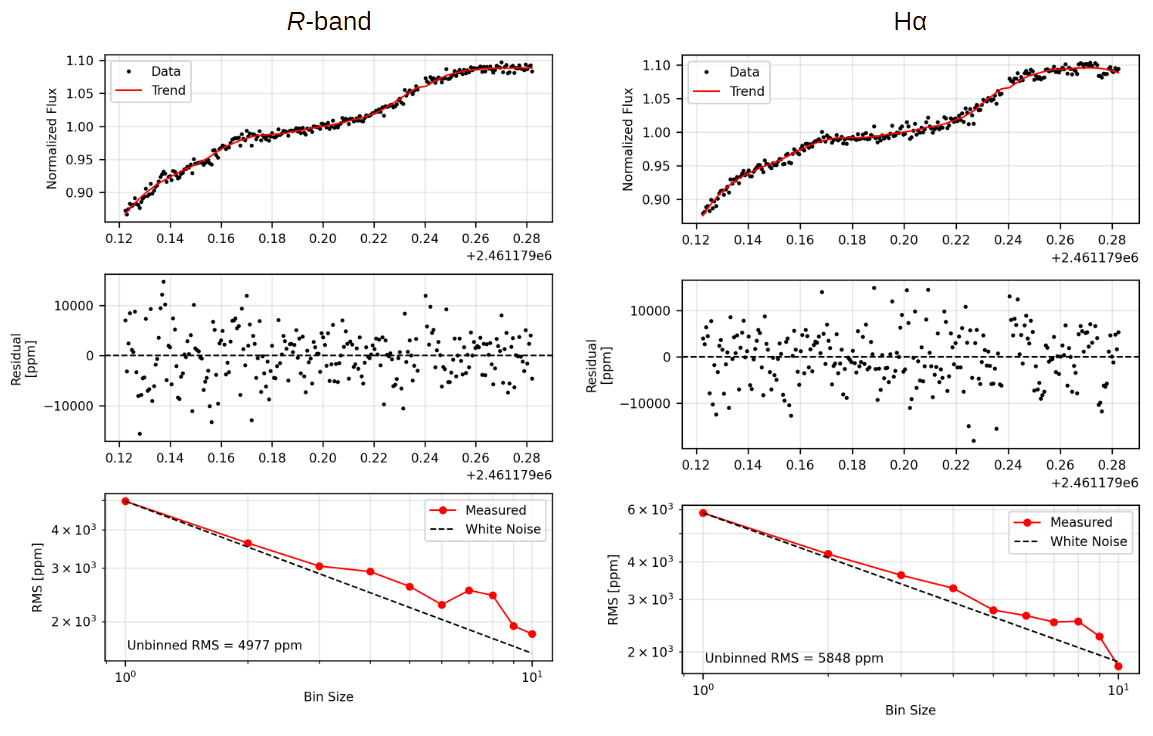}
    \end{tabular}
  \end{center}
  \caption 
  { \label{fig:precision}
  Representative photometric stability measurements in the $R$ (left) and
  H$\alpha$ (right) bands. Top: normalized light curves and
  Savitzky--Golay trends. Middle: detrended residuals. Bottom: RMS as a
  function of bin size; dashed lines show the white-noise expectation. } 
\end{figure} 

First, the absolute photometric precision achieved in each filter was evaluated using the HD~189733 observations. 
To remove astrophysical variability and long-timescale atmospheric fluctuations, a third-order Savitzky--Golay filter was applied to each light curve. 
The filter window length was set to 60 min to isolate the short-timescale
noise relevant to instrumental performance. This detrending was used only
for the precision analysis and not for the transit fits described below.

The photometric precision was estimated from the root-mean-square (RMS) scatter of the residuals after detrending. 
In addition, the RMS was calculated after binning the data by factors of 2, 4, and 8 in order to investigate the timescale dependence of the noise and assess the presence of time-correlated (red) noise.

The resulting photometric precisions are summarized in Table~\ref{tab:precision}. 
Representative examples of the detrending procedure and residuals for the $R$ and H$\alpha$ bands are shown in Figure~\ref{fig:precision}.
The unbinned RMS is lowest in Na D (0.34\%) and highest in $B$
(0.94\%), broadly following the detected counts in
Table~\ref{tab:sequence}. After binning eight consecutive measurements, the
RMS decreases to 0.16--0.40\%. The decrease is slower than the
$N^{-1/2}$ relation expected for purely white noise, particularly in the
broadband light curves, indicating a residual time-correlated component.

The measured unbinned RMS exceeds the expected statistical noise in
Table~\ref{tab:sequence} by factors of 5.17, 1.90, 4.07, 1.79, and 2.54 in
$R$, Na D, $V$, H$\alpha$, and $B$, respectively. Thus, the narrow-band
measurements obtained with 20-s exposures approach within a factor of two
of the source-, sky-, and read-noise expectation, whereas the 1-s
broadband exposures show a larger excess. This difference is qualitatively
consistent with short-exposure scintillation and other rapid atmospheric
fluctuations, which are averaged more effectively during the longer
narrow-band integrations. The ISAS telescope is located on a rooftop at
the Sagamihara campus. Temperature differences between the roof and ambient
air, heat released by the building, and airflow around nearby structures
may generate local optical turbulence and thereby increase seeing-related
photometric variations.
Because no contemporaneous turbulence profile or scintillation monitor was
available, the individual atmospheric contributions cannot be separated
quantitatively. Nevertheless, the comparison shows that the achieved
precision is within a factor of a few of the statistical limit under the
commissioning conditions.

\subsubsection{Relative Photometric Precision}
\label{sec:relative_precision}
\begin{table*}
\centering
\caption{
Photometric precision of the filter-ratio light curves and inferred
correlation coefficients between each filter pair.
The correlation coefficient was estimated from the measured
RMS values assuming
$\sigma_{A/B}^2=\sigma_A^2+\sigma_B^2-2\rho\sigma_A\sigma_B$.
}
\label{tab:relative}
\begin{tabular}{|l|cccc|cccc|}
\hline
&
\multicolumn{4}{c|}{RMS [\%]} &
\multicolumn{4}{c|}{Correlation coefficient $\rho$} \\
\cline{2-9}
Filter Ratio &
Bin 1 &
Bin 2 &
Bin 4 &
Bin 8 &
Bin 1 &
Bin 2 &
Bin 4 &
Bin 8 \\
\hline
Na D / {\it R}
& 0.52 & 0.38 & 0.25 & 0.16
& 0.28 & 0.29 & 0.51 & 0.57 \\
{\it V} / {\it R}
& 0.63 & 0.45 & 0.29 & 0.19
& 0.25 & 0.31 & 0.62 & 0.50 \\
H$\alpha$ / {\it R}
& 0.70 & 0.53 & 0.35 & 0.20
& 0.16 & 0.15 & 0.53 & 0.68 \\
{\it B} / {\it R}
& 1.04 & 0.78 & 0.57 & 0.35
& 0.06 & 0.10 & 0.22 & 0.30 \\
\hline
\end{tabular}
\end{table*}
Next, we evaluated the performance of relative photometry between different filter bands. 
Relative fluxes were calculated by averaging the measurements obtained within each filter-switching cycle (approximately one minute) and taking the ratio between the corresponding bands. 
The photometric precision of the resulting ratio light curves was quantified using the RMS of the detrended residuals, and the results are summarized in Table~\ref{tab:relative}.

To assess the degree of common-mode noise shared between filters, we estimated an effective correlation coefficient, $\rho$, from the measured RMS values under the assumption
\begin{equation}
\sigma_{A/B}^{2} = \sigma_{A}^{2} + \sigma_{B}^{2} - 2\rho\sigma_{A}\sigma_{B},
\end{equation}

where $\sigma_{A}$ and $\sigma_{B}$ denote the photometric scatter in the individual bands and $\sigma_{A/B}$ represents the scatter of the flux ratio. 
For small fractional variations, the residual of $A/B$ is approximately
$\delta_A-\delta_B$. Variations with the same sign in both bands therefore
cancel in the ratio. Consequently, a positive $\rho$ reduces
$\sigma_{A/B}$ below the value $\sqrt{\sigma_A^2+\sigma_B^2}$ expected for
independent variations; the continuum band can thus serve as a reference
for suppressing variations shared with the narrow band.
The inferred correlation coefficients are also listed in Table~\ref{tab:relative}.

All filter pairs exhibit positive correlations, indicating that a fraction of the observed noise originates from common-mode sources such as atmospheric transparency variations, image-motion effects, and long-term instrumental drifts. 
The strongest correlations are found for the narrow-band to continuum combinations, particularly H$\alpha$/$R$ and Na D/$R$. 
While the inferred correlation coefficients are modest for the unbinned data ($\rho \sim 0.1$--$0.3$), they increase substantially with bin size and reach $\rho \sim 0.5$--$0.7$ for eight-point bins. 
This behavior suggests that short-timescale noise components are largely uncorrelated between filters, whereas longer-timescale fluctuations are shared among bands and can be effectively suppressed through ratio photometry.

The highest correlation is observed for the H$\alpha$/$R$ ratio, reaching $\rho = 0.68$ at a bin size of eight measurements, while Na D/$R$ exhibits $\rho = 0.57$. 
The reduction of the ratio RMS associated with these positive correlations
shows that the filter-switching approach implemented in DOLPHIN removes a
fraction of the low-frequency variation shared by the narrow and continuum
bands on timescales of several minutes. 
Since exoplanet transit and atmospheric absorption signals typically evolve on timescales of hours, the observed increase in correlation with bin size suggests that the technique is well suited for detecting such long-duration signals from ground-based observations.

\subsubsection{End-to-End Signal Recovery}
As a final system-level check, the $V$- and $R$-band differential light
curves were fitted with a standard transit model. The recovered
planet-to-star radius ratios, $0.157\pm0.012$ in $V$ and
$0.170\pm0.008$ in $R$, agree with the adopted literature value of 0.155
within $2\sigma$. This test confirms that the complete acquisition and
reduction chain preserves a known broadband signal. Simplified fits to the
filter-ratio light curves, with the transit geometry fixed, yielded
differential radius ratios of $0.005\pm0.005$, $0.010\pm0.004$,
$0.021\pm0.005$, and $0.024\pm0.007$ for $V/R$, Na D/$R$, H$\alpha$/$R$, and $B/R$, respectively. The
H$\alpha$/$R$ fit provided the largest improvement over the null model
($\Delta\chi^2_{\rm null}=16.5$), whereas $V/R$ was consistent with zero.
Because the ratio curves retain correlated structure and the filters were
sampled sequentially, these exploratory results demonstrate signal-recovery
capability but are not interpreted as planetary-atmosphere detections.
Detailed modeling is presented in Appendix~\ref{app:signal_recovery}.

\section{Summary}

We developed DOLPHIN, a compact photometric instrument that rapidly switches
between broadband continuum filters and narrow-band filters centered on Na
D and H$\alpha$. Its single-channel architecture places all bands on the
same detector and optical path, while an eight-position motorized wheel
switches adjacent filters in approximately 0.8~s. This approach trades
strict simultaneity for a simple, reproducible, and comparatively
inexpensive instrument suitable for 1-m-class telescopes.

Laboratory characterization verified the principal requirements for
precision photometry. The detector response is linear to within 0.1\% below
10,000~ADU. We measured a conversion factor of 0.81~e$^-$/ADU, read noise
of 9~e$^-$, and pixel-response non-uniformity of 0.37\%. The dark current
is approximately 0.01~e$^-$ per pixel per second. The measured Na D and
H$\alpha$ passbands are centered at 589.227 and 656.379~nm, respectively.
The H$\alpha$ FWHM is 1.177~nm, whereas the Na D FWHM of 2.232~nm is
broader than its catalog specification; this difference should be included
in future forward models of expected absorption signals.

Commissioning observations of HD~189733b demonstrate the on-sky performance
of the complete system. The unbinned single-band RMS ranges from 0.34\%
for Na D to 0.94\% for $B$ and decreases to 0.16--0.40\% after
eight-point binning. Filter-ratio light curves show increasing positive
correlation with bin size, reaching $\rho=0.68$ for H$\alpha$/$R$ and
$\rho=0.57$ for Na D/$R$, which demonstrates suppression of low-frequency
common-mode variations. The recovered broadband transit provides an
end-to-end check of the acquisition and reduction chain. The ordering of
the measured precision among filters broadly follows the available photon
counts, and the achieved scatter is within a factor of a few of the
statistical expectation after atmospheric and instrumental contributions
are considered. The present work therefore shows that scientifically useful
differential narrow-band time-series measurements are feasible with a
1-m-class telescope. DOLPHIN provides a practical route for
modest-aperture facilities to participate in spectral-line monitoring
programs that have commonly relied on substantially larger telescopes.

\appendix
\renewcommand{\theHsection}{appendix.\Alph{section}}
\renewcommand{\theHsubsection}{appendix.\Alph{section}.\arabic{subsection}}
\renewcommand{\theHsubsubsection}{appendix.\Alph{section}.\arabic{subsection}.\arabic{subsubsection}}

\section{On-Sky Operational Details}
\label{app:operation}

\begin{figure}
  \begin{center}
    \begin{tabular}{c}
    \includegraphics[width=0.95\linewidth]{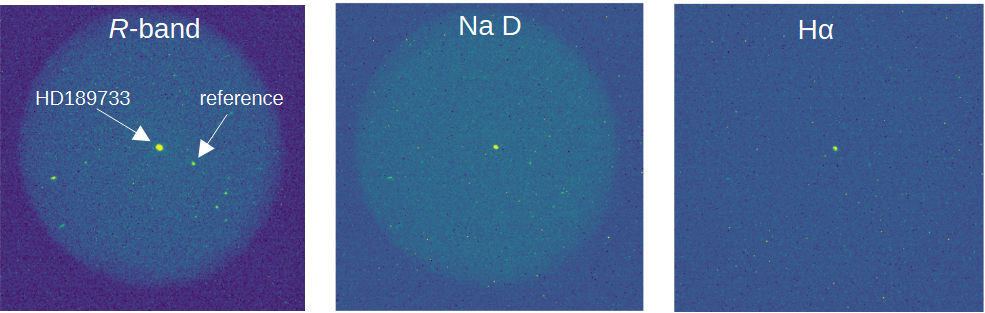}
    \end{tabular}
  \end{center}
  \caption{\label{fig:target}
  DOLPHIN $V$-band image of the HD~189733 field. The target and the
  $V=10.9$ comparison star used for differential broadband photometry are
  marked.}
\end{figure}

To maintain accurate target positioning, the centroid of HD~189733 was
measured after each exposure. Telescope pointing corrections were applied
when the centroid deviation exceeded 5 pixels. Given the DOLPHIN plate scale
of 0.2 arcsec pixel$^{-1}$, this threshold corresponds to approximately
1 arcsec on the sky. The typical site seeing is greater than 2 arcsec. To avoid
corrections driven by chromatic image shifts, guiding feedback was enabled
only for the $V$- and $R$-band images.

The secondary-mirror position was adjusted for each filter to compensate
for chromatic focus offsets. The filter order was selected to minimize
focus and guiding adjustments. HD~189733 was placed near the detector
center, within the nominal filter-response region, while retaining the
$V=10.9$ comparison star shown in Figure~\ref{fig:target} within the field.
No comparison star of comparable brightness to HD~189733 was available in
the approximately 10-arcmin field of view.

\section{Photometric Reduction}
\label{app:reduction}

Each science frame was calibrated with master dark and flat frames. For
each source, the centroid was estimated and pixels within 100 pixels of the
centroid were examined. The source aperture was defined as the pixels whose
signals exceeded the local median by more than $0.5\sigma$, where $\sigma$
was estimated from the median absolute deviation.

The sky level was measured in an annulus beginning 10 pixels outside the
source region. The net source count was
\begin{equation}
\label{eq:net_source}
F = \sum_{i \in {\rm aperture}} S_i-N_{\rm ap}\bar{B},
\end{equation}
where $S_i$ is the signal in pixel $i$, $N_{\rm ap}$ is the aperture area,
and $\bar{B}$ is the mean sky level. The associated photometric uncertainty
was estimated as
\begin{equation}
\sigma_F=\sqrt{F+N_{\rm ap}\sigma_B^2},
\end{equation}
where $\sigma_B$ is the standard deviation of the background pixels in
electron units.

\section{Astrophysical Signal-Recovery Test}
\label{app:signal_recovery}

HD~189733b has previously shown wavelength-dependent Na D and H$\alpha$
transit signatures in high-resolution spectroscopy.\cite{Redfield2008,Jensen2012}
Here these bands are used only for an exploratory end-to-end test of
DOLPHIN; the analysis is not intended as an independent atmospheric
detection.

\begin{figure}
  \begin{center}
    \begin{tabular}{c}
    \includegraphics[width=13.5cm]{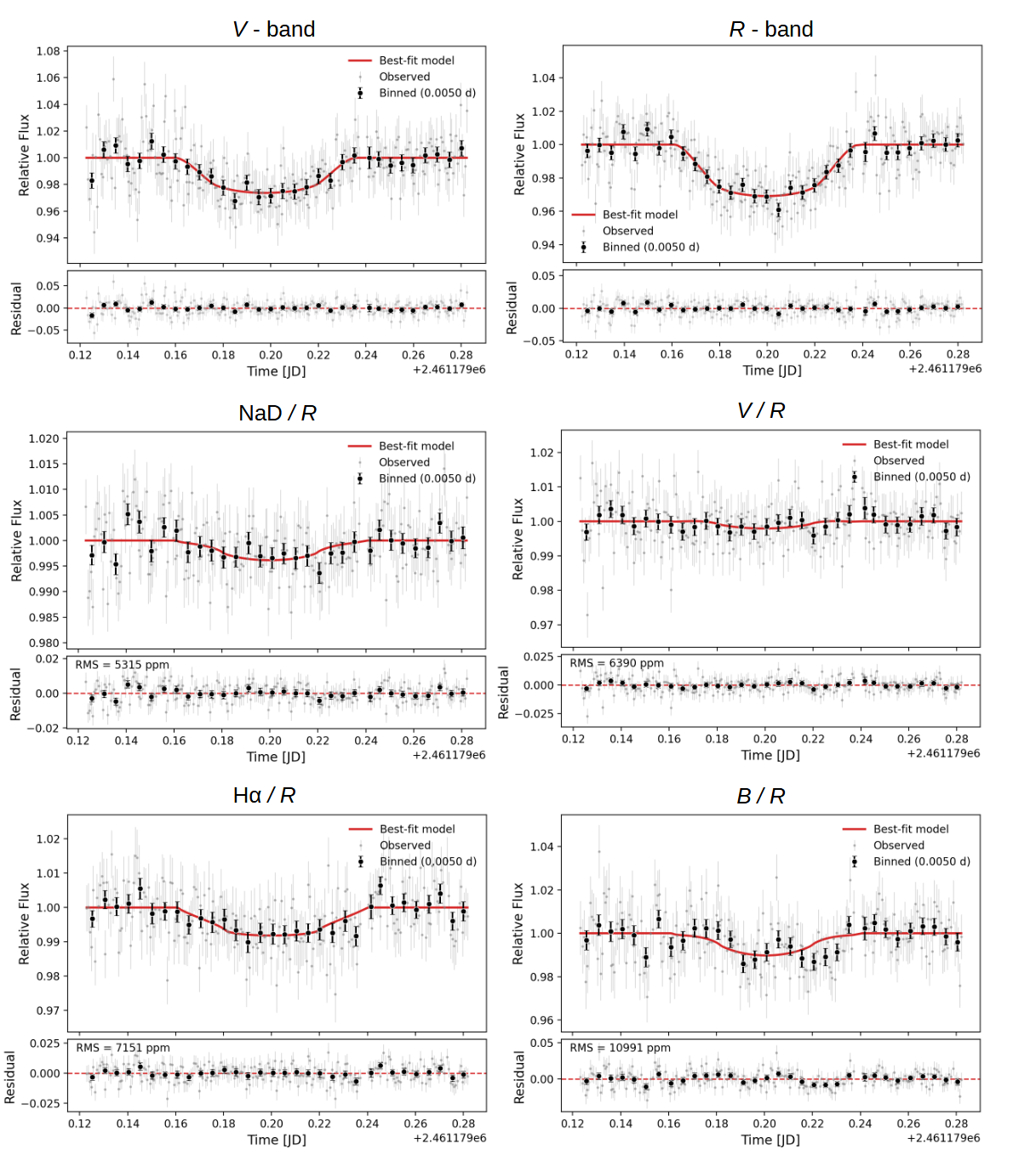}
    \end{tabular}
  \end{center}
  \caption 
  { \label{fig:transitfit}
  Recovered transit signals of HD~189733b. The top panels show the
  differential $V$- and $R$-band light curves, and the remaining panels
  show the Na D/$R$, $V$/$R$, H$\alpha$/$R$, and $B$/$R$ ratio light
  curves. Gray points represent the individual measurements, black points
  show data binned in 0.0050-day intervals, and the red curves are the
  best-fit models. The residuals are shown below each light curve. } 
\end{figure} 

\begin{table}
\centering
\caption{\label{tab:stellarparams}
Stellar and planetary parameters of HD 189733 adopted in this work.
}
\begin{tabular}{|lcc|}
\hline
Parameter & Value & Citation \\
\hline
$T_{\rm eff}$ [K] & $4875\pm43$ & \cite{Boyajian2015} \\
$R_s$ [$R_\odot$] & $0.805\pm0.016$ & \cite{Boyajian2015} \\
$\log g$ & $4.56\pm0.03$ & \cite{Boyajian2015} \\
$[{\rm Fe/H}]$ & $-0.03\pm0.08$ & \cite{Boyajian2015} \\
\hline
$T_c$ [JD] & 2454279.436 & \cite{Agol2010} \\
$P$ [days] & 2.21857567 & \cite{Agol2010} \\
$a/R_s$ & 8.863 & \cite{Agol2010} \\
$i$ [deg] & 85.71 & \cite{Agol2010} \\
$R_p/R_s$ & 0.155 & \cite{Agol2010} \\

\hline
\end{tabular}
\end{table}

\begin{table}
\centering
\caption{\label{tab:transitfit}
Transit fit results in $V$ - and $R$ - bands for HD 189733 b.
}
\begin{tabular}{|c|cc|}
\hline
  &
$V$ &
$R$ \\
\hline

$R_p/R_s$ & $0.157_{-0.012}^{+0.012}$ & $0.170_{-0.008}^{+0.008}$\\
\hline
\end{tabular}
\end{table}

\begin{table}
\centering
\caption{\label{tab:absorption}
Differential planet-to-star radius ratios measured in the $V$, Na D,
H$\alpha$, and $B$ bands relative to the $R$-band continuum. The second row
gives the improvement in $\chi^2$ relative to a model with
$\Delta r_p=0$.
}
\begin{tabular}{|c|cccc|}
\hline
& $V/R$ &
${\rm NaD}/R$ &
${\rm H\alpha}/R$ &
$B/R$ \\
\hline
$\Delta r_p$ & $0.005_{-0.005}^{+0.005}$  & $0.010_{-0.004}^{+0.004}$ & $0.021_{-0.005}^{+0.005}$& $0.024_{-0.007}^{+0.007}$ \\
$\Delta \chi^2_{\rm null}$& 7.0 & 6.5 & 16.5 & 11.1 \\

\hline
\end{tabular}
\end{table}
To evaluate the recovery of transit and wavelength-dependent signals with
DOLPHIN, we first analyzed the $V$- and $R$-band differential light curves
using the $V=10.9$ comparison star. The stellar and planetary parameters
adopted in the analysis are summarized in Table~\ref{tab:stellarparams}. We
generated transit light curves with \texttt{batman}\cite{Kreidberg2015} and
adopted a quadratic limb-darkening law. The coefficients for each passband
were fixed to values calculated with \texttt{LDTk}\cite{Parviainen2015}
using the stellar parameters of HD~189733. We sampled the posterior
distribution with the affine-invariant Markov chain Monte Carlo sampler
\texttt{emcee}.\cite{Foreman-Mackey2013} The transit-center time,
planet-to-star radius ratio, baseline flux, and an additive jitter term were
allowed to vary; the remaining orbital parameters were fixed to literature
values.\cite{Agol2010} To account for excess scatter beyond the formal
photometric uncertainties, the log-likelihood included the jitter term $s$:
\begin{equation}
\ln \mathcal{L} = -\frac{1}{2}\sum_{i=1}^{N}
\left[
  \frac{\left(y_i - m_i\right)^2}{\sigma_i^2 + s^2} +
  \ln\left\{2\pi\left(\sigma_i^2+s^2\right)\right\}
\right].
\end{equation}
where $y_i$ is the observed flux, $m_i$ is the model flux, $\sigma_i$ is the photometric uncertainty, and $s$ is the jitter term.

The resulting best-fit models are shown in Figure~\ref{fig:transitfit}, and the derived parameters are summarized in Table~\ref{tab:transitfit}.
The transit morphology is recovered independently in both broadband light
curves despite the relatively large scatter of the individual exposures.
The fitted radius ratio in the $V$ band,
$R_p/R_s=0.157\pm0.012$, is consistent with the adopted literature value of
0.155. The $R$-band value, $0.170\pm0.008$, is larger by 0.015, but the
difference is less than $2\sigma$ and therefore does not constitute
significant evidence for a wavelength-dependent broadband radius.  This
agreement demonstrates that the observing sequence and reduction procedure
can recover a transit depth at the few-percent level in independent filters.

We next searched for wavelength-dependent absorption using the ratio light
curves. These were fitted with a ratio of two ordinary transit models rather
than with a single transit model. For a numerator band $X$, the model was
\begin{equation}
M_{X/R}(t)=C_{X/R}
\frac{T_X\!\left[t; r_R+\Delta r_p,\boldsymbol{u}_X\right]}
     {T_R\!\left[t; r_R,\boldsymbol{u}_R\right]},
\qquad r_R\equiv(R_p/R_s)_R,
\end{equation}
where $T_X$ and $T_R$ are \texttt{batman} transit models, $C_{X/R}$ is the
out-of-transit normalization, and $\boldsymbol{u}_X$ and
$\boldsymbol{u}_R$ are the quadratic limb-darkening coefficients calculated
with \texttt{LDTk} for the respective passbands. The orbital geometry and
transit-center time were fixed to the broadband solution, and only
$C_{X/R}$, $\Delta r_p$, and the jitter term were fitted. Thus,
$\Delta r_p=0$ defines the null ratio model. The results are listed
in Table~\ref{tab:absorption}.  The $V/R$ result,
$\Delta r_p=0.005\pm0.005$, is consistent with zero, as expected from the
similar broadband transit depths. In contrast, the H$\alpha$/$R$ ratio gives the
largest improvement over a null model, with
$\Delta\chi^2_{\rm null}=16.5$, followed by $B/R$ with 11.1.
The Na D/$R$ improvement is weaker
($\Delta\chi^2_{\rm null}=6.5$).  Figure~\ref{fig:transitfit} shows that the
corresponding depressions occur over the same time interval as the
broadband transit. This temporal agreement is consistent with an
astrophysical signal, but does not by itself exclude correlated systematics.

These apparent excesses should nevertheless be interpreted as a
demonstration of signal recovery rather than as definitive detections of
the planetary atmosphere.  The inferred differential radii,
$0.010\pm0.004$ for Na D, $0.021\pm0.005$ for H$\alpha$, and
$0.024\pm0.007$ for ${\rm B}$, are comparable to or smaller than the scatter of
individual ratio measurements, and the residuals retain structure on
timescales comparable to ingress and egress.  Moreover, the alternating
filter sequence samples the bands non-simultaneously.  Changes in
atmospheric transparency, seeing, image position, and airmass within a
filter cycle can therefore produce chromatic residuals that are not fully
removed by division by the $R$ band.  Stellar activity and differences in
limb darkening between the line and continuum bands may also mimic or alter
a differential transit depth.

Nevertheless, the recovery of transit-shaped signals in all four
filter-ratio light curves is consistent with the common-mode suppression
inferred in Section~\ref{sec:relative_precision}. H$\alpha$/$R$ gives the
largest $\Delta\chi^2_{\rm null}$, whereas Na D/$R$ has the lowest
ratio-light-curve RMS among the tested pairs but a smaller improvement over
its null model. Additional transits,
comparison with out-of-transit control sequences, and a joint model of
airmass and instrumental systematics will be required to establish the
repeatability and astrophysical origin of these excess signals.  The present
result therefore validates DOLPHIN's ability to recover broadband transits
and demonstrates its potential for differential narrow-band
spectrophotometry.

\subsection*{Disclosures}
The authors declare no conflicts of interest.

\subsection*{Acknowledgments}
This work was supported by Astrobiology Center PROJECT research, Grant Number AB0517 and JSPS KAKENHI grant Nos.26H02072, 26K00752, and 26H02074 (H.K.). 
The data analysis used the \texttt{batman}, \texttt{LDTk}, and \texttt{emcee}
software packages. OpenAI Codex was used to assist with English-language
editing and the organization of the manuscript. The author reviewed all
AI-assisted revisions and takes full responsibility for the scientific
content, interpretation, and conclusions.

%
\bibliography{report}
\bibliographystyle{spiejour}
%
%
%
%

\end{document}